\documentclass[runningheads]{llncs}
\usepackage[T1]{fontenc}
\usepackage{color}
\usepackage{graphicx}
\usepackage{caption}
\usepackage{subcaption}
\usepackage{amsmath}
\begin{document}
\title{Analysis of Reinforcement Learning for determining task replication in workflows}
\titlerunning{Reinforcement Learning for determining task replication in workflows}
%
\author{Andrew Stephen McGough\and
Matthew Forshaw}
\authorrunning{McGough and Forshaw}
%
\institute{School of Computing, Newcastle University, UK
\email{\{stephen.mcgough,matthew.forshaw\}@ncl.ac.uk}}
\maketitle              
\begin{abstract}
Executing workflows on volunteer computing resources where individual tasks may be forced to relinquish their resource for the resource's primary use leads to unpredictability and often significantly increases execution time. Task replication is one approach that can ameliorate this challenge. This comes at the expense of a potentially significant increase in system load and energy consumption. We propose the use of Reinforcement Learning (RL) such that a system may `learn' the `best' number of replicas to run to increase the number of workflows which complete promptly whilst minimising the additional workload on the system when replicas are not beneficial. We show, through simulation, that we can save 34\% of the energy consumption using RL compared to a fixed number of replicas with only a 4\% decrease in workflows achieving a pre-defined overhead bound.
\end{abstract}
%
%
%
\section{Introduction}
Workflows comprising several independent (computational) tasks under a strict ordering (see Figure~\ref{simpleGraph} for an example) have become one of the pillars of computational research and industrial development. Users wish to enact these as quickly as possible -- within a factor of the critical path execution time -- referred to as the contingency. However, many organisations lack dedicated resources, nor the operational expense for Cloud enactment for this work. Here these users are required to `make do' with computational resources which are not dedicated to this purpose -- often referred to as volunteer computing. One of the common ways of providing this is through High Throughput Computing (HTC), which exploits the idle time available on computing resources provisioned for other purposes -- examples of such systems include HTCondor~\cite{tannenbaum2001condor} and BOINC~\cite{boinc}.

Although HTC systems allow workload to be performed on volunteer resources -- thus sharing one of the main advantages of Cloud computing of having no (or marginal) capital expense -- it does have the distinct disadvantage that resources can (and often are) retracted without warning for their primary use. This will lead to the termination of those tasks which are currently being executed on that resource. Approaches to ameliorate this include suspension of tasks until the resource is available again~\cite{suspend}, relaunching the task on a new resource to run from the beginning~\cite{reducing} or checkpointing and migration~\cite{niu2013employing}. These approaches ensure eventual task completion, but impact the overall execution time of the task~\cite{mcgough2013ccpshort} -- which can be compounded when enacting a workflow.

Users will, in general, expect their workflows to finish in a ``reasonable'' time, relative to execution time for the workflow. Given that the tasks within the workflow will often have to relinquish the resource back to the primary user there can be considerable variance to the workflow execution time (see Section \ref{psa}). Leading to users believing that their workflow has crashed or the system is unfit for purpose. However, this could just be a consequence of `bad luck' for the particular workflow enactment. Ideally a workflows execution time should be relative to the length of the workflow's critical path (defined as the shortest time from the start to the end of the workflow~\cite{criticalPath}) where a subset of the workflow tasks form the critical path. We formally state the users desire here that a given workflow will complete within time $(1+p)CR(W),$ where the delay proportion $p \ge 0$ and $CR(W)$ is the time to execute the critical path for workflow $W$.

Although a task may not initially form part of the critical path if it is forced to relinquish a resource the delay to its completion may make it part of a newly revised critical path. As such, the workflow enactment needs either to reduce the chance of the task from becoming so delayed or compensate for such a delay. 

Task replication -- running multiple copies of the same task -- can ameliorate the impact of such delays. However, running these extra tasks can quickly impact other work on the cluster leading to all workflows taking longer to complete and higher overall energy consumption -- as all but one are wasted work. Determining how many replicas to have is a difficult problem and is related to the resources primary usage pattern, how critical the task is to timely completion and other workloads on the system. We propose using Reinforcement Learning (RL)~\cite{rl} to `learn' the number of replicas to deploy based on the time of day and how critical a tasks is to the workflow finishing within its contingency requirements.

Section \ref{psa} discusses the problem space and analyse the impact of running workflows in non-dedicated environments. Related work is evaluated in Section \ref{back}. Our approach is presented in Section \ref{method}. Section \ref{experiment} presents our simulation model before presenting results in Section \ref{results}. We present our conclusions in Section \ref{conc}.

\section{Problem Space Analysis} \label{psa}
\begin{figure}[t]
\label{graph}
  \begin{center}
  \includegraphics[width=6cm]{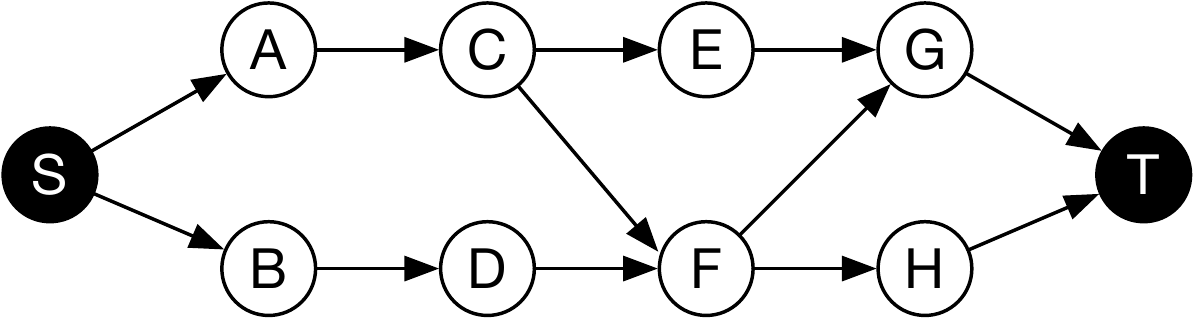}
  \caption{A workflow. Nodes are tasks to run and links represent dependencies.}
  \label{simpleGraph}
  \end{center}
\end{figure}

We define, without loss of generality, a workflow as a Directed Acyclic Graph (DAG) comprising independent tasks (represented as nodes) with an associated ordering on their execution (represented by directed links between nodes). Tasks have an associated execution time, including time for setup, data ingress and egress, whilst links have associated time required for scheduling. A DAG may have more than one start/end node. Without compromising our model, we assume that such DAGs have an extra `task' at the start (end), providing a single start (end) node. These tasks have zero execution time; however, the link execution time may be non-zero -- to cater for workflow setup or termination. Figure \ref{simpleGraph} illustrates a simple DAG. The normal tasks (A -- H) have links to indicate their dependencies (e.g. task F can only execute once both tasks C and D have completed). Extra tasks are added to give a single start (S) and termination (T).

We analyse the impact on a workflow when running it on volunteer resources such as a University HTC system. We extend HTC-Sim~\cite{htc-sim} for workflows~\cite{workflow}. As the original trace-log does not contain workflows we replace 10\% of the original tasks in the trace-log with synthetic workflows (Figure \ref{simpleGraph} with each task taking 32 mins). For each workflow which completed we compute the value of $p(W)$, the proportion of excess time to execute the workflow, as:
\begin{equation}
    p(W) = \frac{(c(W)-s(W))}{CR(W)} - 1
    \label{pp}
\end{equation}
where $c(W)$, $s(W)$ are the finish and submission times of tasks T, S from W and $CR(W)$ is the execution time for the critical path of $W$.
\begin{figure}[b!]
\centering
\begin{subfigure}{.5\textwidth}
  \centering
  \includegraphics[width=113pt,height=113pt]{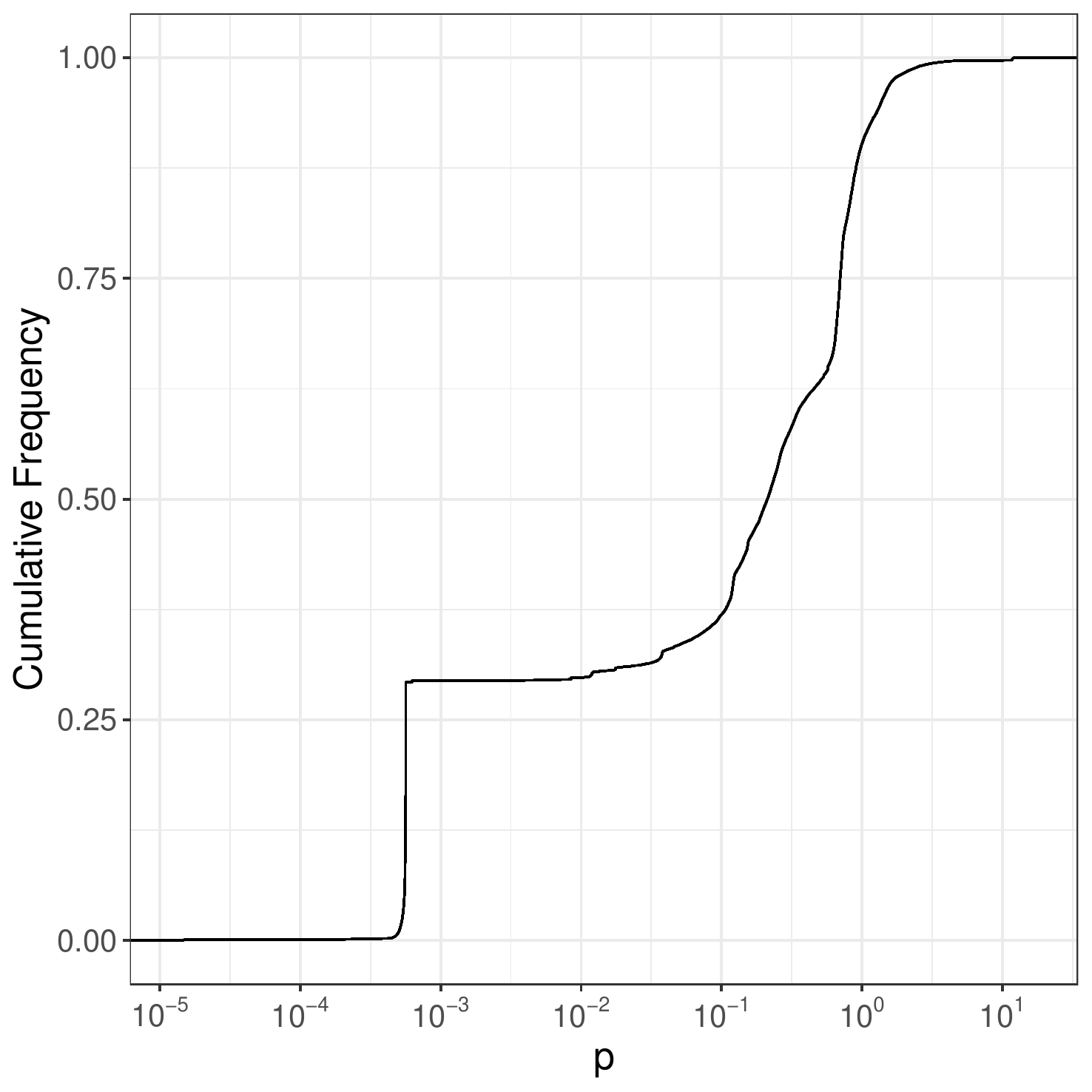}
  \caption{CDF for $p$}
  \label{cdf}
\end{subfigure}%
\begin{subfigure}{.5\textwidth}
  \centering
  \includegraphics[width=.743\textwidth]{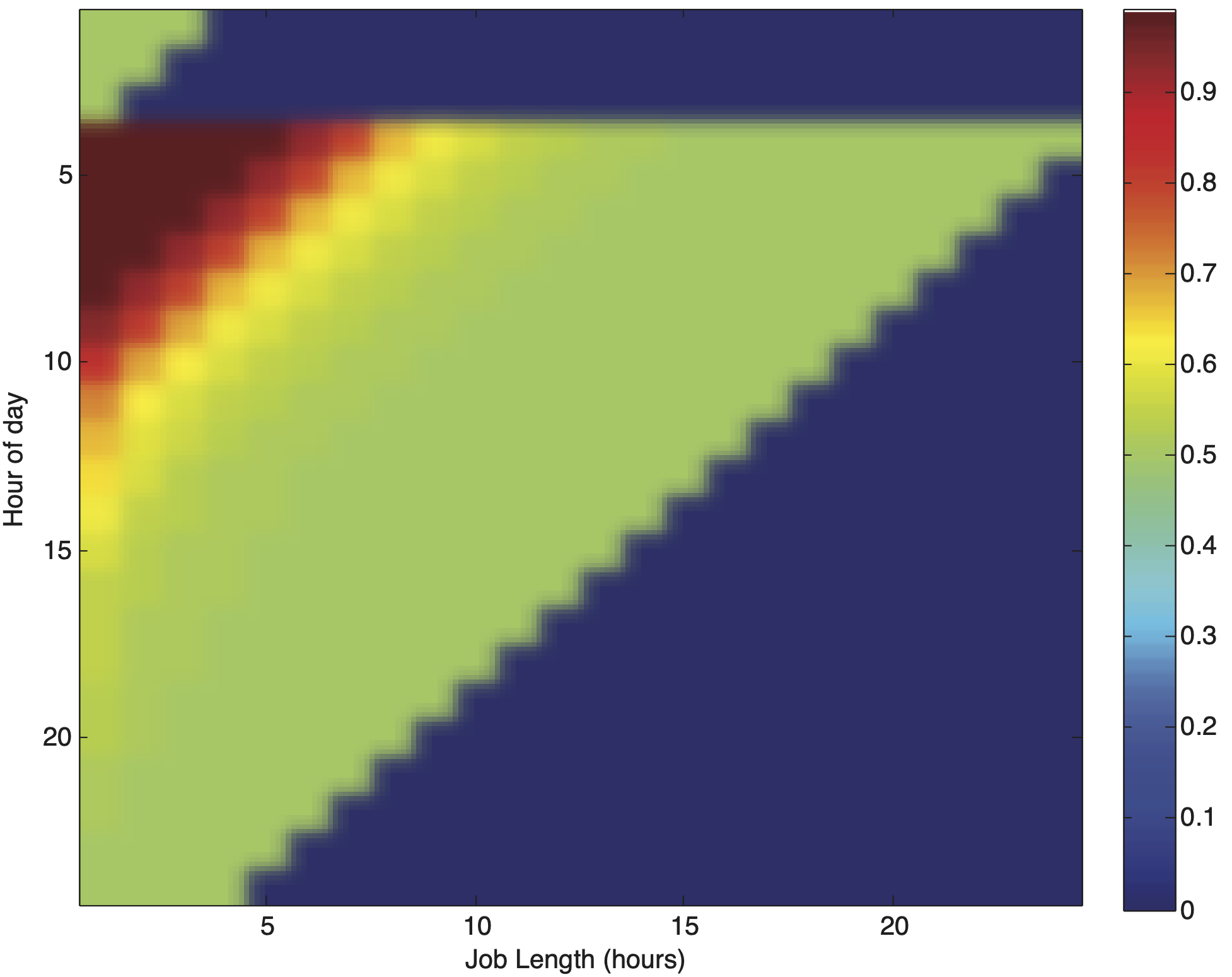}
  \caption{Probability of task completing based on length and time of submission}
  \label{heat}
\end{subfigure}
\caption{CDF and probability of task completion}
\label{heatmap}
\end{figure}

Simple statistical analysis (mean 0.4467, standard deviation 0.8217) of the 40,054 workflows suggests little impact from running in a volunteer environment. However, the Cumulative Density function (Figure \ref{cdf}) illustrates the long-tail on these results. Very few workflows complete with a value of p(W) less than 0.00004. With a significant ($\sim30\%$) of workflows having $p(W)\sim0.000043$. On the other hand, 10\% of the workflows take double or more than their critical path time to run ($p(W)\geq1$). These would seem to match those situations where the system was under exceptional load causing workflow tasks to be repeatedly forced to relinquish resources. Our worst case was when $p(W)=17.1113$.

The probability of each individual task in a workflow completing given its length and time of submission is presented in Figure \ref{heat}. Short tasks submitted early in the morning have the highest probability of success with reasonable chances across most of the day. The discontinuity at 3am and the reduction to zero probability for longer jobs later in the day is a consequence of the nightly reboot at 3am. For this system we predict that replicas are likely to be beneficial through the day, but less so in the early morning or before the 3am reboot. In this work we propose an adaptive system which `learns' system-specific behaviours.

\section{Related Work} \label{back}

\subsection{Energy-aware workflow scheduling}
The challenges of scheduling workflows to computational resources has received extensive treatment in the literature. A detailed exposition of previous efforts is presented by Yu {\em et 
al.}~\cite{yu2008workflow}. Existing approaches predominantly focus on workflow scheduling performance, with comparatively few considering the energy consumption as a primary optimisation goal. Durillo~\emph{et al.}~\cite{durillo2013multi} adapt the well-established HEFT~\cite{topcuoglu2002performance} heuristic for workflow scheduling, and demonstrate the potential for significant energy conservation, for a small performance penalty.

Several approaches explore enacting workflows in the Cloud.
Zhu~\emph{et al.}~\cite{zhu2010power} propose \emph{pSciMapper} for energy-aware consolidation of workflow tasks in virtualised environments. 
Li~\emph{et al.}~\cite{li2016security} adopt a particle swarm optimisation (PSO) approach to workflow scheduling in the Cloud, with consideration for financial cost and security. Calheiros~\emph{et al.}~\cite{calheiros2014meeting} focus on task replication to overcome issues introduced by performance variation of public Cloud resources.


Our approach is distinct from those presented. Firstly, we consider workflow enactments to non-dedicated resources shared with their primary users. Secondly, we adopt a RL based approach to learn characteristics of the environment to which the workflows are deployed.

\subsection{Simulations with support for workflows}
The benefits of using simulation frameworks to evaluate the performance of scheduling policies in large-scale computing systems is well understood~\cite{forshaw2015operating,IoTCase}. There exist a number of simulators for Grid and Cluster computing infrastructures~\cite{optorsim,gridsim,simgrid}, these lack modelling capabilities to evaluate the energy consumption of the simulated infrastructure. Simulation tools including~\cite{cloudsim,kliazovich,mdcsim,sicogrid} have the ability to evaluate energy-performance. However, these tools do not allow modelling of tasks needing to relinquish resources and workflows. With the exception of SimGrid, which is capable of modelling DAG-based workflows, these tools lack the capability to model dependencies between jobs. 
Chen and Deelman~\cite{chen2012workflowsim} extend CloudSim~\cite{cloudsim} to model Workflows, and evaluate the performance of several common workflow scheduling heuristics.

Previously we extended HTC-Sim~\cite{forshaw2014trace}, a Java-based trace-driven HTC simulation, to support workflow execution~\cite{workflow}. This enhancement will be used here, allowing us to explore the energy and performance impact of workflow enactment schemes within a multi-use cluster environment.

\section{Reinforcement Replication} \label{method}

We present our replication approach within a volunteer HTC environment, however, it is equally applicable to other environments exhibiting intermittent faults.

\subsection{Approach overview}

We augment an event based model~\cite{Georgakopoulos1995} for workload enactment to support task replication, in which every time the status of the workflow changes (e.g. workflow starts, task completes) an event is triggered and the appropriate actions are performed. We assume here that a user has defined $\phi$\footnote{Where the user desires $p(W) \leq \phi$} their contingency proportion for workflow execution. Our adaptations to these workflow event handlers are:
\begin{itemize}
	\item {\bf Workflow starts:} The desired deadline $d$ for the workflow is computed:
	\begin{equation}
	    d = (\phi+1)CR(W) + s(W).
	\end{equation}
	Next, for each task $t$ within the workflow which can be run -- effectively those tasks linked to task `S' -- we determine their local $\phi'(t)$\footnote{To simplify presentation we omit the parameters hereafter.} -- reflecting how `critical' $t$ is to the timely completion of the workflow and is computed by removing those tasks from $W$ which do not depend on $t$ (to give $W'$) and computing the new critical path ($CR(W')$). Then $\phi'$ can be computed using equation~\ref{pp} with $\phi' = p(W)$, $s(W) = s_t$ (the submission time of task $t$), $c(W) = d$ and $CR(W) = CR(W')$. Here $\phi'$ will be the proportional difference between the critical path and the time to the deadline $d$. Note that $\phi' \geq \phi$ with equality when $t$ is part of the critical path for the whole workflow. The number of replicas to be run can now be determined through the use of RL which is parameterised by both the time of the day at which task $t$ is submitted and the value of $\phi'$.
	\item {\bf Task completes:} We define task completion to be the time at which the first replica of a given task completes. First all other replicas are terminated as they are no longer of any use. We can then determine any new tasks for the workflow which are now free to be executed -- those task for which all prior task have completed. For each of these tasks $t$ we can compute $\phi'$ (as above) and hence the number of replicas which should be deployed at this time of day, until the workflow has completed.
\end{itemize}

\subsection{Metrics}

In order to evaluate the effectiveness of our approach and to provide metrics for use by our approach we define the following metrics:

\begin{itemize}
	\item {\bf Number of workflows with an overall $p(W)\leq P$}: Our users seek to have their workflows finish within the shortest time which is proportional to the length of the critical path of the workflow. We seek to identify the number of workflows which complete with a contingency less than $P$.
	\item {\bf Energy consumption from replicas}: Here we report both the `good' energy (that was used for the successful task) and the `bad' energy for replica tasks. In the case of `bad' energy this is the combination of both replicas which fail to be the first to complete and any replica runs which are terminated before completion. More formally for task $i$ this is defined as:
		\begin{equation}
		\label{bad}
		bad_i = \sum_{r \in X_i} \sum_{j \in I_{i,r}} \left\{ 
  \begin{array}{l l}
    \tau_{i,r,j} E_{i,r,j} & \quad \textrm{if } G_{i,r,j} \neq 1\\
    0       & \quad \textrm{otherwise}\\
  \end{array} \right. ,
		\end{equation}
		where $X_i$ is the set of all replicas of the task $i$, $I_{i,r}$ is the set of all invocations of replica $r$, $\tau_{i,r,j}$ is the execution time of replica $r$ invocation $j$, $E_{i,r,j}$ is the energy consumption rate (Watts) of the resource selected for replica $r$ invocation $j$ and $G_{i,r,j}$ is one for the good replica invocation, else zero. Likewise:
		$$
		good_i = \tau_{i,r,j} E_{i,r,j},\ \ \  r \in X_i, j \in I_{i,r}, \text{ such that } G_{i,r,j} = 1.
		$$
		Note that only one $G_{i,r,j} = 1$. Note that the total energy consumed for running the replicated task will be $good + bad$.
\end{itemize}

\subsection{Computation of $\phi'$}
\label{criticalP}

In order to compute $\phi'$ we must first determine the time remaining before we reach the workflow deadline and the estimated time that the rest of the workflow will take to complete. The first value can be computed as $d - s_t$. The second value, the time to complete task $t$ and all those tasks which are dependant on $t$, is computed as:
\begin{equation} \label{critPath}
 L_t =
  \begin{cases}
    r_t + \max_{j \in C_t} L_j & \quad \text{if $|C_t| \neq 0$} \\
    r_t		     				     & \quad \text{otherwise}\\
  \end{cases},
\end{equation}
where $C_t$ is the set of tasks imediately dependant on task $t$, and $r_t$ is the duration of task $t$. We consider $L_t$ to be the local critical path from task $t$ to the end of the workflow. Note that for simplicity here we assume $r_t$ includes any time for data ingress / egress and setup time. Note also that if $t$ is the start task of the workflow then $L_S$ is the  duration of the critical path for the whole workflow. 

Given that before workflow execution starts we do not know the values of $r_t$ we can replace these with estimates ($e_t$) of the execution time. At this stage we do not concern ourselves as to how these estimates are derived except to note that this could be provided through a performance repository used to predict task execution times~\cite{escPredict} or the use of performance prediction~\cite{predictable}. Equation \ref{critPath} becomes:
\begin{equation} \label{critPathQoS}
 L'_{t} =
  \begin{cases}
    e_t + \max_{j \in C_t} L'_{j} & \quad \text{if $|C_t| \neq 0$} \\
    e_t 		         & \quad \text{otherwise}\\
  \end{cases}.
\end{equation}

The workflow will rarely follow this expected execution duration. Tasks may take more or less time than their estimate -- either due to miss-estimates of their execution time or due to the fact that multiple submissions are required before certain tasks complete. This can lead to tasks, which weren't on the critical path, becoming part of the new critical path. Before submitting each task within the workflow we need to determine its potential impact on the overall workflow and how we can minimise this impact -- say by running multiple copies of the task to increase its chance of completing within its own expectation time ($(1+\phi)e_t$).

We have evaluated two approaches can be taken for determining the potential impact of a task on the overall workflow -- both based around $\phi$:

\subsubsection{All remaining tasks balanced:}
In this case we compute the spare time available for all task (including this task) which depend on this task and divide this proportionally between all these tasks. Thus we compute $\phi'$ as the proportion we can allocate to all remaining tasks that are dependant on task $t$:
$$
\phi' = \frac{d-s_t}{L'_{t}} - 1.
$$
With three cases for the value of $\phi'$:
\begin{itemize}
  \item {\bf $\phi \leq \phi'$}: Here, for this path of the workflow we are currently not part of the critical path, or we are on the critical path but ahead of schedule. There is currently no additional risk to completing the workflow by $d$. Running replicas of task $t$ can be performed if deemed necessary by the time of day.
  \item {\bf $0 \leq \phi' < \phi$}: We have fallen behind schedule for this path of the workflow, though there is still a probability that the workflow can be finished successfully. Remedial actions can be taken (such as running multiple copies of the current task) to increase the chance of completing the workflow by $d$.
  \item {\bf $\phi' < 0$}: The workflow is significantly behind schedule and this path is likely to prevent the workflow from completing before $d$, remedial work is required. Note that if the remaining tasks in this path complete faster than their expected execution time it may still be possible to finish by $d$.
\end{itemize}

\subsubsection{Balance on current task:}
Instead of balancing all remaining slack time across the remaining tasks in this path we assume all remaining tasks will still have their proportion $\phi$ while the current task will do anything possible to get itself back on track. We define $\phi''$ as the proportion of extra time available to task $t$:
$$
\phi'' = \frac{d - s_t - \max_{j \in C_t} L'_{j}}{e_t} - 1.
$$
The three cases for $\phi'$ above also apply to $\phi''$, though for $\phi'' < 0$ there is still slack time for the remaining tasks so the workflow may still finish on time. 

\subsection{Reinforcement Learning}
We wish to determine for a given task $t$, submitted at time $s_t$, and with contingency proportion $\phi'$\footnote{without loss of generality we use $\phi'$ to represent both $\phi'$, and $\phi''$.} the number of replicas which should be submitted. We include here one as a valid number of replicas. In order to tailor our approach to an individual HTC system we use Reinforcement Learning (RL)~\cite{rl}, to train an agent to provide the number of replicas which is expected to give the greatest reward (chance that the task will complete in the minimal time). RL is a form of Machine Learning which can learn the `best' action to perform given a particular system state. This can be achieved without training data -- with training coming from the interaction between an agent and a reward function which provides feedback on the actions taken by the agent. Thus RL can, not only, adapt itself to any given environment but also, as it continually trains, adapt as the environment changes. RL has been previously used to solve control problems such as elevator scheduling, resource allocation within a data centre~\cite{mlDC}, reduction of energy consumption within volunteer computer systems~\cite{suscom} and bidding stratergies for energy markets~\cite{kell}.

In order to use RL to optimise the number of replicas we use the approach of an n-armed bandit~\cite{rl}. Under this assumption each action -- the number of replicas to run -- is independent of all other actions performed. 

Each task $t \in \{1, 2, ...\}$ which we wish to replicate will observe the system in a given state $s \in S$. Our state space here represents those characteristics of the system over which decisions should be made -- in this case the time of day at which the task is to be submitted and the contingency proportion available for the task. As these need to be discrete values we round the time of submission to the hour of the day when submission took place, thus giving 24 states for time. Likewise for contingency proportion we discretise this to $n+2$ intervals. Namely:
$$
\{\phi' \leq 0, 0 < \phi' \leq \frac{P}{n}  , ..., \frac{P(n-1)}{n} < \phi' \leq P , P < \phi' \}.
$$
Thus our state space comprises of $24(n+2)$ states. In order to maintain our n-armed bandit model we assume that a task which has to relinquish a resource becomes a new task within the system when it is re-allocated. The set of actions ($a \in A$) is the number of replicas which to be submitted to the system. We determine the action $a$ to perform as:
\begin{equation}
a = f(Q(s,A)),
\end{equation}
where $Q(s,A)$ is the set of all reward values for the actions $A$ available when the system is in state $s$ and $f()$ is a selection policy. The true reward values $Q(s,A)$ are unknown, but we can estimate $Q'(s,A)$ from the prior decisions which have been made and the associated rewards. This becomes an estimator for $Q(s,A)$:
\begin{equation}
Q'_{t}(s,A) = \{q'_{t}(s,a) \} \quad \forall a \in A,
\end{equation} 
\begin{equation}
q'_{t}(s,a) = \overline{R_i(s,a')} \quad \forall i \leq t, a' = a,
\label{average}
\end{equation}
\noindent where $R_t(s,a) \in [-k,k]$ is the reward for taking action $a$ in state $s$ for task $t$. A value, for $R_t(s,a)$, of $-k$ indicates this was the worst possible choice of action whilst $+k$ indicates the best choice of action. The value of $k$ can be chosen arbitrarily, however, it is normally a small number to prevent buffer overflows. There are two possible outcomes when an action is applied to a task. These are:
\begin{itemize}
\item {\bf Task $t$ completed within the contingency proportion $\phi'$:} This is seen as a success for action $a$ which should be rewarded thus increasing the chance of this action being selected again in the future. However, if left unchecked using just a reward for success here could lead to a system learning that the highest reward is obtained by allocating the maximum number of replicas to each task. We therefore need to diminish this reward proportional to the wasted work completed for those replica tasks which have been run.
\item {\bf Task $t$ failed to complete within the contingency proportion $\phi'$:} This is seen as a failure for the action $a$ chosen and requires a punishment to reduce the chances of this action being selected again in the future.
\end{itemize}

\noindent Therefore we can define the reward function as follows for task $t$:
\begin{equation*}
 R_t(s,a) = \left\{ 
  \begin{array}{l l}
    +1-\sigma_t & \quad ${$t$ completed within $\phi'e_t$}$\\
    -5 & \quad ${$t$ failed to complete within $\phi'e_t$}$\\    
  \end{array} \right.,
  \label{eqn3}
\end{equation*}
where the first term in the reward function is used to indicate that the chosen action was good or not and the second term (if present) helps to steer the replication task towards the minimum value. We set the value for failed tasks to -5 to incur a large penalty for failure, the RL approach was not significantly afected by changing this value. We set the value of $\sigma_t \in [0, 1]$ to be proportional to the wasted work performed by replicas. We compute $\sigma_t$ using the energy consumption of the `bad' replicas as a proxy for the wasted work:
$$
\sigma_t = \min (1, \frac{bad_t}{a d_t \Xi}),
$$
where $\Xi$ is the average energy consumption rate for the selected resource when performing work -- we assume 100\% utilisation.

We can now define the selection policy $f()$ which is used to evaluate the action to perform given the prior history reward set $Q'(s,A)$. We define two approaches, a greedy (exploitative) and an explorative selection policy:
\begin{equation}
f(Q'(s,A)) =  \left\{ 
  \begin{array}{l l}
    max_a(Q'(s,A)) & \quad ${with probability 1 - $\epsilon$}$ \quad ${(exploitative)}$\\
    random(A)        & \quad ${with probability $\epsilon$}$ \;\; \quad \quad ${(explorative)}$\\
  \end{array} \right.
\end{equation}
Here $max_a()$ selects the action $a$ with the highest expected reward, whilst $random(A)$ will select an action uniformly at random from $A$.

By selecting the greedy policy we are exploiting prior knowledge to use the action with the greatest expected reward, whilst an exploitive policy allows us to search for potentially better actions. The dynamic and changing nature of our system necessitate both exploitative and explorative policies. Being too greedy can lead to poor choice of replica counts as the agent will keep using sub-optimal actions, whilst being too explorative can lead to the use of sub-optimal actions which are known to be bad. A careful selection of $\epsilon$ is therefore required.

\subsubsection{Vary $\epsilon$:}
In most cases, like ours here, the RL system will start off in an uninitiated state where each action has the same reward value. If the value of $\epsilon$ is too small then the system can keep choosing the wrong action, when in exploitation mode, due to insufficient training. Likewise, if the underlying system changes the `learned' actions may no longer be valid -- in which case we should return to an explorative policy. We can therefore choose to vary $\epsilon$ during execution to allow better training. Two common approaches are:

\begin{itemize}
\item {\bf Initially high $\epsilon$:} Initially the value of $\epsilon$ is set high ($\epsilon_1$), then to a lower value $\epsilon_2$ after the first $l$ rewards have been observed. This allows the RL to be initially more explorative and once the system has had a chance to `learn' the best actions it will revert to a more exploitative policy.

\item {\bf Vary $\epsilon$ when results of choosing an action vary from those expected:} Here a sliding window captures the results from the last $m$ selections of a given action. If the average reward of this value deviates too far then the value of $\epsilon$ can be increased for a time until it is deemed that the system has been re-trained. This allows the RL to become more explorative when the rewards move away from the expected range, to adapt to underlying changes. 
\end{itemize}

\subsection{Mechanisms of workflow task enactment}

In this work we compare three mechanisms for workflow task enactment:

\begin{itemize}
	\item {\bf Single task execution:} Each task within a workflow is submitted only once to the HTC system. This provides a baseline for workflow execution times.
	\item {\bf Fixed replica execution:} The number of replicas is fixed. Although this can lead to reduced execution time of the workflow it has two main disadvantages. Firstly, tasks will be needlessly replicated at times of the day when this is not required. Secondly, the extra replicas could lead to contention for limited resources, hindering tasks which otherwise would have completed.
	\item {\bf RL-based replica execution:} The number of replicas is determined dynamically at runtime. The likelihood of workflows finishing within the defined contingency is increased, whilst minimising overheads and contention.
\end{itemize}

\section{Experimental Setup} \label{experiment}

We extend the HTC-Sim simulation~\cite{htc-sim,workflow} to model task replicas and incorporate RL functionality. We use our trace logs~\cite{htc-sim} from the use of the HTCondor~\cite{htcondor} system and interactive users at Newcastle University during 2010. These logs represent 1,229,820 interactive user logins -- the primary users of the cluster of ~1,400 computers. During the year a total of 561,851 HTC tasks were submitted. As our original trace-log lacks examples of workflows, we randomly replace w\% of the tasks within the trace-log with synthetic workflows. Here we choose w = 10\% and use the same modified trace-log for all experiments to ensure consistency. We experiment with two workflow types: the simple workflow from Figure~\ref{simpleGraph} (each task taking 32 minutes), more real-world workflows generated by Montage.

We have limited the maximum number of replicas, of each workflow task, to ten as experiments have shown that values greater than this give no advantage and often cause the cluster to become overloaded. Values of $\phi$ range between 0.1 and 1.0 in steps of 0.1 -- we also evaluate the value of $\phi=2$. For our state-space we have chosen $n = 10$ thus matching in with the intervals used for $\phi$. We varied $\epsilon_1 \in [0.05,1]$ and $\epsilon_2 \in [0.05,0.4]$ and the sliding window $m \in [10,10,000]$.

\section{Results} \label{results}

We present here only the results for the ``All remaining tasks balanced'' case for brevity as we observed no statistical difference between the two different approaches. We used $\epsilon_1 = 0.5, \epsilon_2 = 0.05$. Table \ref{base} presents the base case, no task replication, indicating the number of workflows which completed within $P$ contingency (Successful) and the energy consumed in these cases (Energy in MWh). The number of workflows which complete successfully increases with $P$ as expected. Although the values for energy are varying this is more due to the randomness within the simulation (the time the workflows are executed and which resources are used to run them).


\begin{table}[!t]
\centering
  \fontfamily{ppl}\selectfont
\begin{tabular}{c||c c c c c c c c c c c}
$\phi$ & 0.1 & 0.2 & 0.3 & 0.4 & 0.5 & 0.6 & 0.7 & 0.8 & 0.9 & 1.0 & 2.0\\ \hline
Successful & 14,982 & 19,622 & 22,979 & 24,791 & 25,475 & 26,494   & 29,967   & 33,081   & 35,015   & 36,318   & 39,650\\
Energy &  28.83 &  30.86 & 30.87 & 29.86 & 29.08 & 29.78 & 30.98 & 30.25 & 28.92 & 28.49 & 30.72\\
\end{tabular}
\caption{Baseline case}
\label{base}
\end{table}%

\begin{figure}[t]
\begin{center}
  \includegraphics[width=\textwidth]{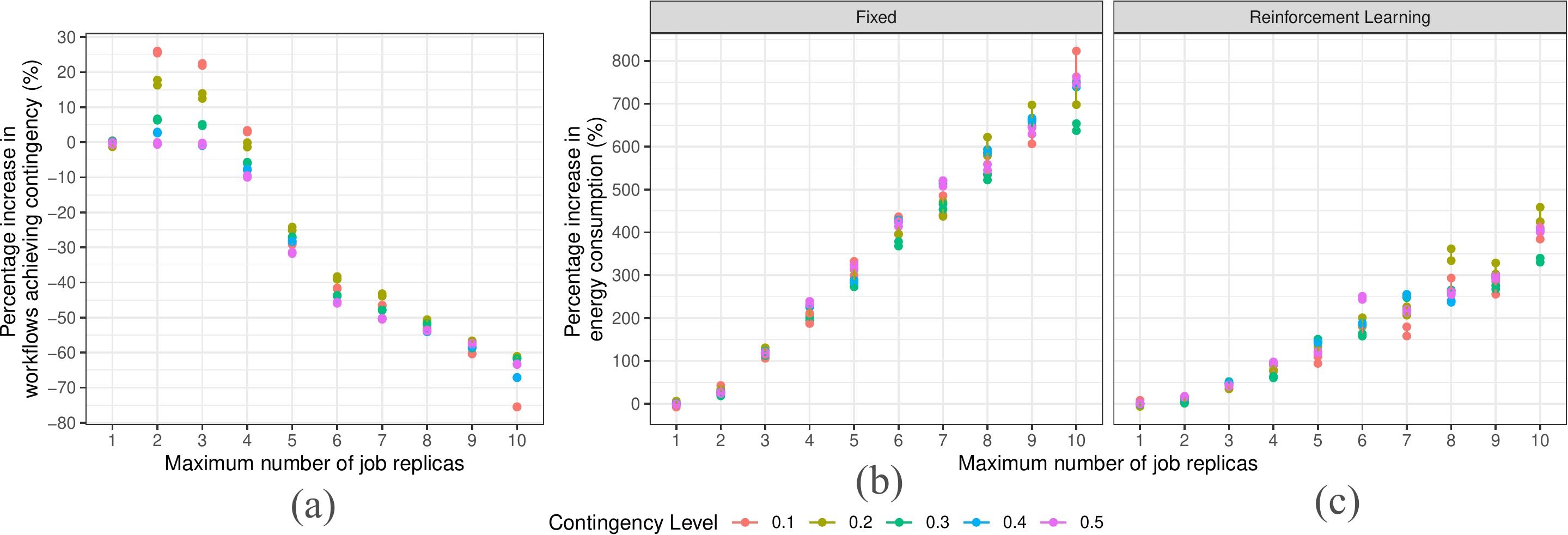}
\end{center}
  \caption{For fixed replicas: a) Percentage of workflows achieving users desired contingency. b) Energy requirements for achieving users desired contingency. c) Percentage increase in energy consumption by using RL.}
  \label{asdf}
\end{figure}

Figure \ref{asdf}a shows the increase in workflows which finish within a contingency proportional to $P$ by using a fixed number of replicas, compared to our baseline case (Table \ref{base}). We see favourable results in the two- and three-replica case. This is particularly evident for contingency value $P$=0.1, where over 20\% more workflows are capable of meeting the target. Four or more replicas lead to contention for resources and corresponding performance degradation. However, if the number of replicas can vary, then it may be beneficial to deploy more than four replicas at specific times of the day or for particular contingency levels.


Figures~\ref{asdf}b and \ref{asdf}c indicate the extra energy required when we run a fixed number and RL chosen replica count for each task within a workflow. As can be seen from this figure the energy consumption increases linearly with the increase in replicas (for the RL case this is the maximum number of replicas it can select). As such this will quickly lead to excessive extra computation and wasted work. However, the RL case effectively halves the energy consumption. 

\begin{figure}[b]
\begin{center}
    \includegraphics[width=0.9\textwidth]{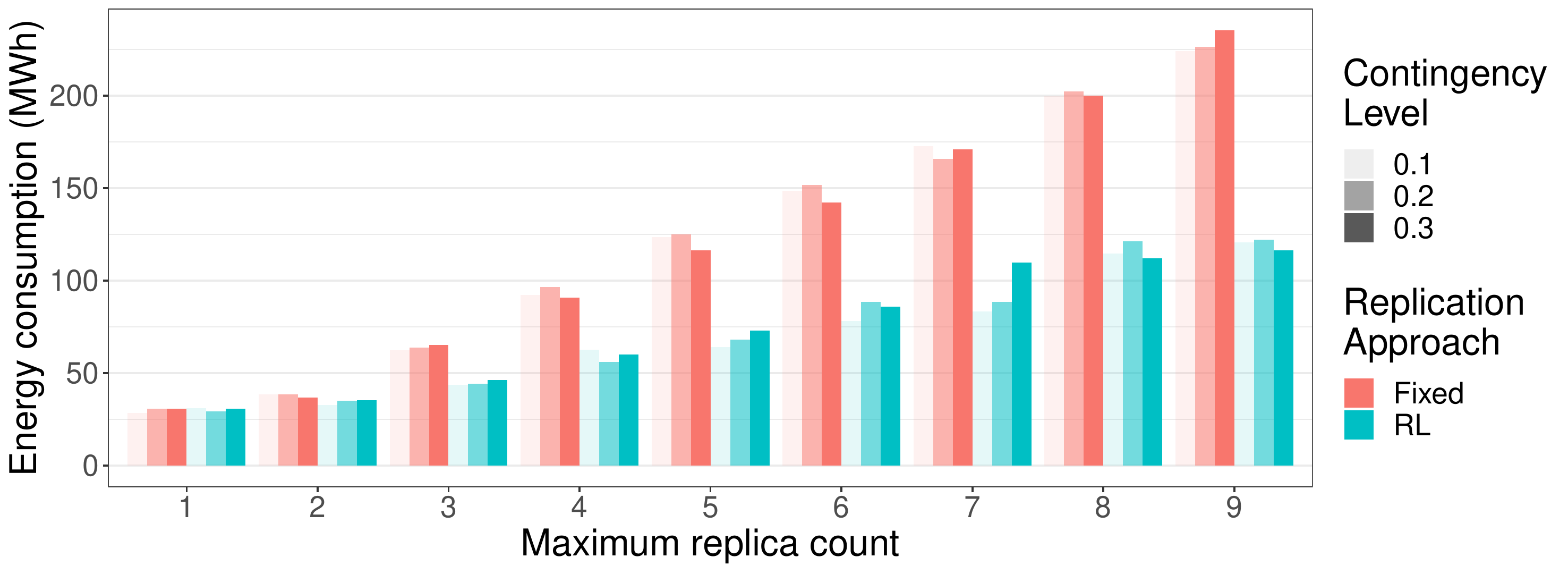}
\end{center}
  \caption{Comparison between energy increase for fixed replicas and RL replicas}
  \label{energyDecrease}
\end{figure}
\begin{figure}[hbt]
\begin{center}
      \includegraphics[width=0.9\textwidth]{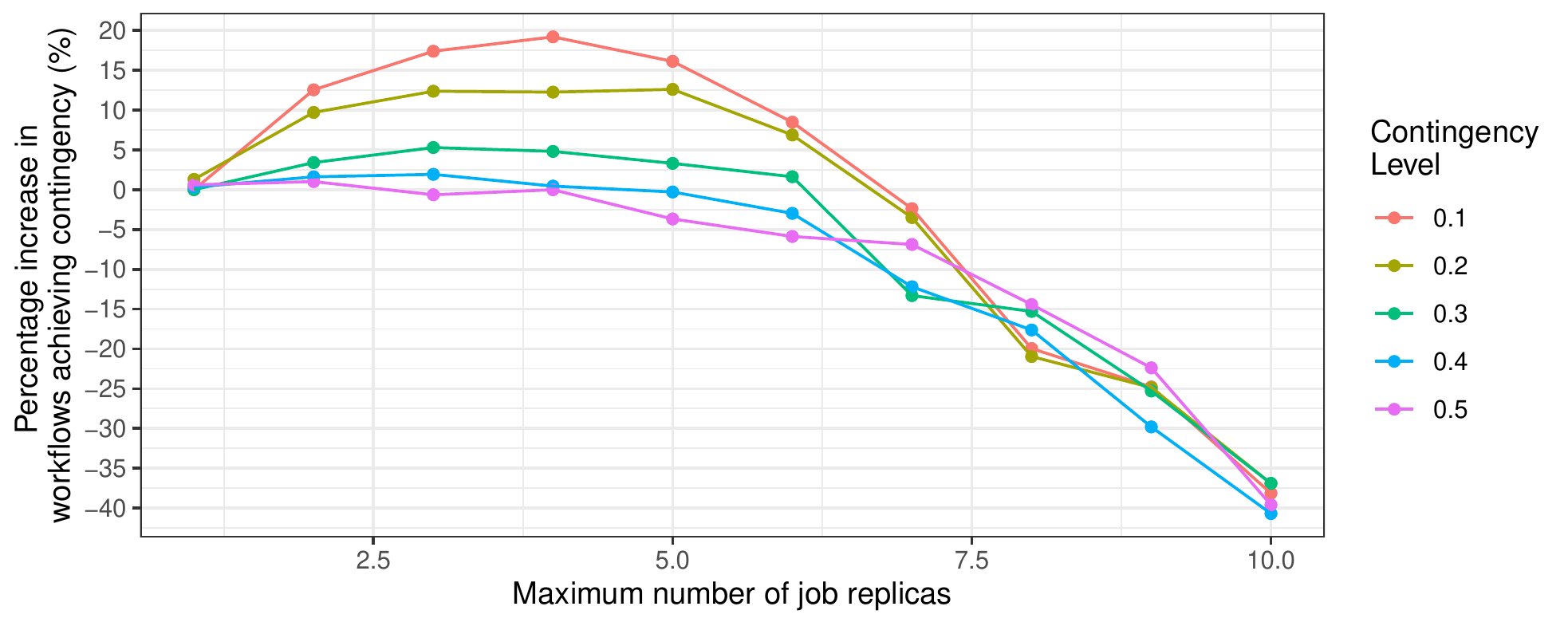}
\end{center}
  \caption{Increase in number of workflows which can be performed when using RL}
  \label{QoSReduce}
\end{figure}

Figure \ref{energyDecrease} shows how much we can reduce the energy consumption of running replicas by allowing RL to select the number of replicas to enact. For fair comparison, RL is bounded by the same maximum number of replicas as the fixed case. The percentage decrease in energy is presented in Figure~\ref{asdf}c. We see our approach achieves promising energy savings for all replication scenarios (replica count $\geq$ 2), successfully identifying system conditions requiring the use of replication, while minimising unnecessary replication during periods of low utilisation which would lead to wasted energy. Although we can significantly reduce the energy consumption of enacting replicas, we must also consider the performance impact of this additional replication. Figure~\ref{QoSReduce} illustrates the proportion of workflows able to complete within time contingency $P$. We observe favourable improvement in workflow completion, particularly when considering small contingency values. Most importantly, we see improvements across a much wider range of replica counts. This shows our approach to be far less susceptible to sub-optimal replica count selection. 


\begin{figure}[hbt]
\begin{center}
      \includegraphics[width=0.85\textwidth]{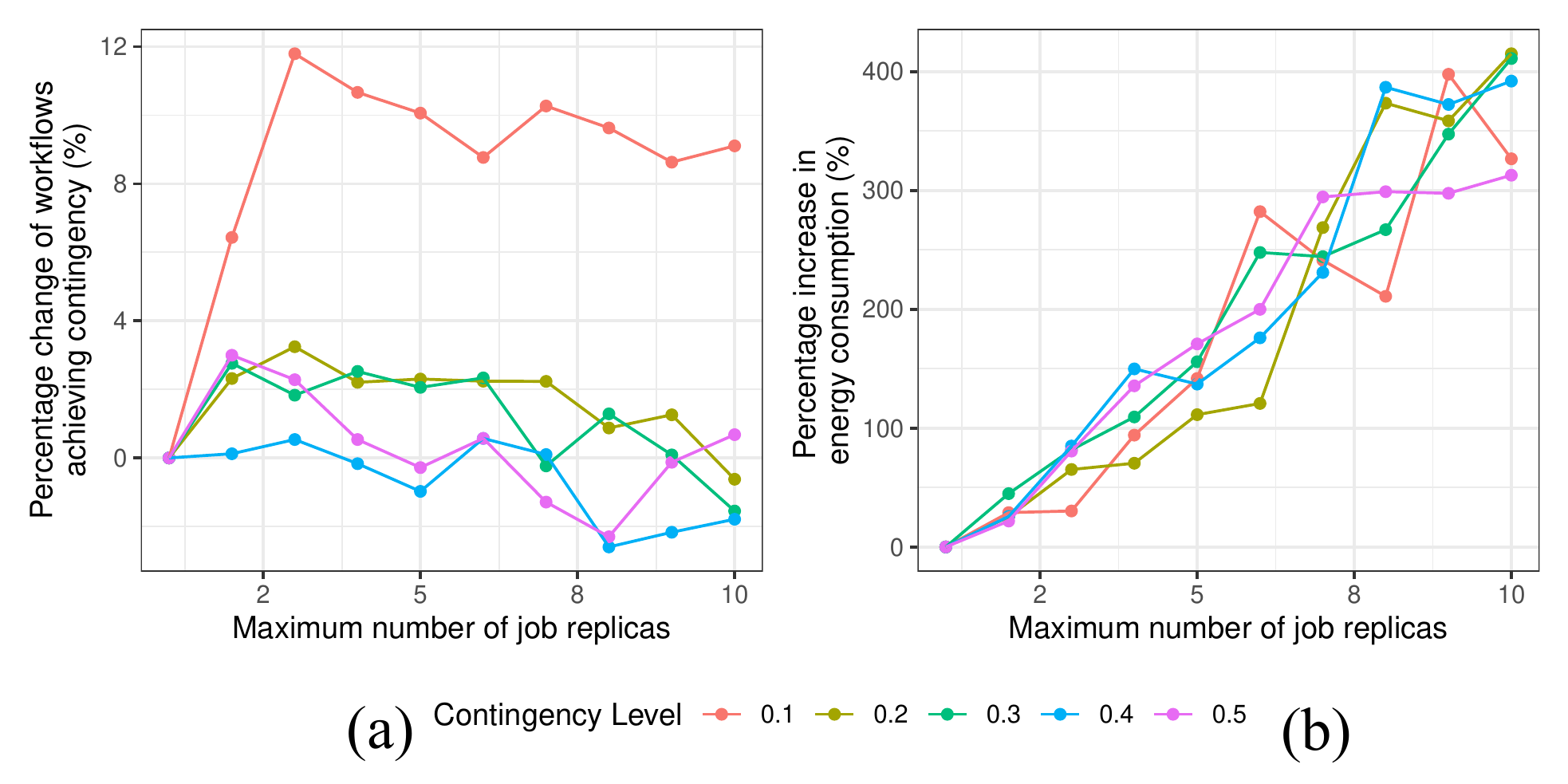}
\end{center}
  \caption{Using RL for Montage. \textbf{a)} Percentage increase in workflows achieving users' desired contingency. \textbf{b)} Percentage change in energy consumption.}
  \label{montage_rl_combo}
\end{figure}
We now validate our approach for the well-established Montage workflow~\cite{jacob2009montage}. Figure~\ref{montage_rl_combo}a shows our approach can improve the number of successful workflows. This seems to be irrespective of the choice of replica count, however, this is most likely a consequence of the newly successful workflows only using a small number of replicas. The biggest impact comes from the choice of contingency -- with lower contingency giving better results. Likely a consequence of RL not over-provisioning for cases where it can't complete on time, thus keeping the system idle for other workflows. Figure~\ref{montage_rl_combo}b confirms that the energy impact is reduced significantly compared to a fixed replication scheme. We are able to conserve 34\% energy consumption with only a 4\% decrease in workflows achieving contingency.

%
%

%
%
%
%
%
%
%
%
%
%
%
%
%
%


\section{Conclusions} \label{conc}

In this paper we have explored, through trace-driven simulation, how the performance and energy consumption of workflow enactments in an HTC environment can be improved through task replication. We demonstrate that fixed replication schemes, if tuned correctly, can deliver significant performance improvements, but impose a considerable overhead in energy consumption. In contrast, our proposed Reinforcement Learning approach curtails the energy consumption while retaining the performance benefits. We show our approach to be less susceptible to poor performance due to sub-optimal replica count selection.

Future work will explore the broader applicability of RL approaches in workflow enactment; e.g. an RL-based overlay to the HEFT scheduling heuristic. 
%
%
\bibliographystyle{splncs04}
\bibliography{epew}
\end{document}